\documentclass[a4paper,10pt,journal]{IEEEtran}

\usepackage{cite}
\usepackage{siunitx}
\DeclareSIUnit \GHz {GHz}
\usepackage{graphicx}
\usepackage{amsfonts}
\usepackage{amsmath}
\usepackage[caption=false,font=footnotesize]{subfig}
\usepackage[hyphens]{url}
\usepackage{algorithm}
\usepackage{algpseudocode}

\newcommand{\ep}{{\mathbb E}}
\newcommand{\pr}{{\mathbb P}}
\newcommand{\vc}[1]{\mathbf{#1}}

\newcommand{\set}[1]{{\mathcal{#1}}}
\newcommand{\FF}[1]{{\mathbb{F}}}
\newcommand{\RRR}{{\mathcal{R}}}

\newcommand{\ind}[1]{\mathbb{I}(#1)}
\usepackage{tikz}
\usetikzlibrary{arrows}

\usepackage{enumerate}

\newtheorem{definition}{Definition}[section]
\newtheorem{lemma}{Lemma}[section]
\newtheorem{theorem}{Theorem}[section]

\newtheorem{remark}{Remark}[section]
\newtheorem{example}{Example}[section]
\newcolumntype{P}[1]{>{\centering\arraybackslash}p{#1}}
\newcolumntype{M}[1]{>{\centering\arraybackslash}m{#1}}
\begin{document}
\title{{Reliability of Broadcast Communications Under\\ Sparse Random Linear Network Coding}}

\author{Suzie Brown, Oliver Johnson and Andrea Tassi
\thanks{
This work is partially supported by the University of Bristol Faculty of Science, the School of Mathematics and the \mbox{VENTURER} Project, which is supported by Innovate UK under Grant Number 102202.

S. Brown and O. Johnson are with the School of Mathematics, University of Bristol, UK (e-mail: {\tt sb13831.2013@my.bristol.ac.uk}, {\tt O.Johnson@bristol.ac.uk}).

A. Tassi is with the Department of Electrical and Electronic
Engineering, University of Bristol, UK (e-mail: {\tt A.Tassi@bristol.ac.uk}).
}}

\maketitle

\begin{abstract}
Ultra-reliable Point-to-Multipoint (PtM) communications are expected to become pivotal in networks offering future dependable services for smart cities. In this regard, sparse Random Linear Network Coding (RLNC) techniques have been widely employed to provide an efficient way to improve the reliability of broadcast and multicast data streams. This paper addresses the pressing concern of providing a tight approximation to the probability of a user recovering a data stream protected by this kind of coding technique. In particular, by exploiting the Stein--Chen method, we provide a novel and general performance framework applicable to any combination of system and service parameters, such as finite field sizes, lengths of the data stream and level of sparsity. The deviation of the proposed approximation from Monte Carlo simulations is negligible, improving significantly on the state of the art performance bounds. 
\end{abstract}


\vspace{-3mm}
\section{Introduction}
{In next-generation networks, reliable broadcast communication is expected to be critical. In particular, this holds true in future networks of self-driving vehicles where road-side base stations (BSs) will broadcast live sensor data~\cite{5G-PPP}. For example in the 5G-PPP's ``bird's eye'' use case, live 3D Light Detection and Ranging (LiDAR) scans of vehicles engaging a traffic junction are broadcast to the incoming vehicles --  enabling them to take an informed decision on how to safely drive through the junction~\cite{5G-PPP}. In this kind of network, a key performance indicator is  the \emph{user delivery probability}, defined as the probability of a user successfully recovering the transmitted data stream.}

Generally, modern communication systems enhance the reliability of Point-to-Multipoint (PtM) data streams by employing Application Level-Forward Error Correction (AL-FEC) techniques, which are usually based on Luby Transform (LT) or Raptor codes~\cite{6416071,7393818}. These kinds of codes only operate to their capacity if large block lengths are employed, which could be a problem in the presence of delay-sensitive services~\cite{6416071}. For this reason, in our system model reliability of PtM data streams is ensured via the Random Linear Network Coding (RLNC) approach~\cite{6994250}.

{The RLNC approach requires the BS to split each PtM data stream into $K$ source packets, which form a \emph{source message}. A sequence of coded packets is obtained in a rateless fashion by linearly combining the source packets. A user recovers the PtM data stream as soon as it collects $K$ linearly independent coded packets~\cite{6994250,EvTassi}. A drawback of the RLNC approach is the computational complexity of the decoding phase, which is a function of $K$ and the finite field size $q$ considered during the encoding phase~\cite{DanielTCOM,ComplexityGE}. Tassi~\textit{et al.}~\cite{7335581} observed that this complexity can be significantly reduced by adopting a \emph{sparse implementation} of the RLNC approach, where the number of non-zero elements in the encoding matrix is smaller. However, as the encoding matrix becomes sparser, the number of coded packet transmissions needed by a user to recover the source message is likely to increase. To date, an exact expression for the user delivery probability as a function of the sparsity and number of coded packet transmissions is still unknown.

The key contribution of this paper is a tight approximation to the user delivery probability in a system where broadcast source messages are protected by sparse RLNC (see Section~\ref{sec.PERF}). Our approximation is valid for any finite field size, sparsity level, and data stream length. As shown in Section~\ref{sec.RES}, the deviation of the proposed  model from simulation results is negligible. Our approximation enables service providers to increase the sparsity level (reducing the complexity of the decoding phase) while ensuring a target user delivery probability.
}

The lack of an exact performance model for sparse RLNC implementations is caused by the lack of an accurate expression for the probability of a sparse random matrix, generated over a finite field, being full rank~\cite{cooper1,7335581,DanielTCOM}. Garrido~\textit{et al.}~\cite{DanielTCOM} proposed models based on absorbing Markov chains to characterize the user performance, in a particular implementation of sparse RLNC where the number of source packets employed to generate each coded packet is \emph{fixed}. This assumption significantly simplifies the performance modeling issue, yet~\cite{DanielTCOM} mostly relies on  Monte Carlo simulation to estimate, via a regression technique, the statistical correlation between the rows of full rank sparse matrices.

{Unlike~\cite{DanielTCOM}, this paper refers to a more general sparse RLNC scheme where a source packet participates in the generation of a coded packet with probability $1-p$, for $0 \leq p \leq 1$. With regards to this general sparse RLNC formulation, Tassi~\textit{et al.}~\cite{7335581} proposed the first performance model valid for any finite field size, data stream length and probability $p$. In particular, the probability bound proved in~\cite[Theorem~6.3]{RSA:RSA1} allowed the authors~\cite{7335581} to derive a tractable but not tight performance bound. More recently, the theoretical framework proposed by A. Khan~\textit{et al.}~\cite{7523985} extended~\cite[Theorem~6.3]{RSA:RSA1} and can be directly used to upper- and lower-bound the user delivery probability. However, again often these bounds are not tight.

In this paper, we address the limitations of the previous studies and provide the following contributions:
\begin{itemize}
\item We propose an accurate expression for the user delivery probability suitable for general sparse RLNC formulations and applicable to any combination of system parameters, which overcomes the lack of generality of the model proposed in~\cite{5978939} and~\cite{DanielTCOM}. In particular, with regards to~\cite{DanielTCOM}, a new set of Monte Carlo simulations are required to re-derive a performance model as the field size, the number of source packets defining a source message or the number of source packets involved in the generation of each coded packet changes.
\item Unlike most recent works~\cite{7335581,7523985} which  build upon~\cite[Theorem~6.3]{RSA:RSA1}, we approximate the user delivery probability by employing a novel mathematical framework based on the Stein-Chein method. In fact, for $K > 10$ and $p \geq 0.7$,~\cite[Theorem~6.3]{RSA:RSA1} notoriously is not a tight lower-bound to the probability of a sparse random matrix being full rank~\cite{7335581}, with a subsequent impact on the estimation of the user delivery probability.
\item Regardless of the field size and level of sparsity of the encoding matrix,  our approximation of the user delivery probability is very close to simulated values. On the other hand, the state of the art upper- and lower-bound to the user delivery probability proposed in~\cite{7523985} significantly deviate from Monte Carlo simulations for a binary field, with the lower-bound performing better than the upper-bound. For larger field sizes, both our approximation and the upper-bound as per~\cite{7523985} to the user delivery probability tightly follow simulation results but, in this case, the lower-bound as per~\cite{7523985} significantly deviates from our Monte Carlo simulations. As such, unlike our approximation, neither the upper-bound nor lower-bound consistently give a tight approximation of the user delivery probability, regardless of the field size.
\end{itemize}}

The rest of the paper is organized as follows. Section~\ref{sec.SM} presents the considered system model. Section~\ref{sec.PERF} discusses the proposed performance characterization model for sparse RLNC implementations and states our novel approximate result in Theorem \ref{th1}. The accuracy of the proposed performance model is considered using Monte Carlo simulation in Section~\ref{sec.RES}. Finally, in Section~\ref{sec.CL}, we draw our conclusions.

\vspace{-2mm}
\section{System Model} \label{sec.SM}
%As shown in Fig.~\ref{fig.f_0},
We consider a system model where one transmitter broadcasts a stream of coded packets to multiple receiving nodes, over a channel with packet error probability equal to $\epsilon$.
We assume that the transmission time of a coded packet is equal to one time step, and that the time needed to transmit $N$ coded packets is equal to $N$ time steps. 

% \begin{figure}[tb]
% \centering
% 	\includegraphics[width=0.5\columnwidth]{fig0.eps}\vspace{-3mm}
% \caption{Considered system model including a flow-chart description of the coded packet generation.}\vspace{-3mm}
% \label{fig.f_0}
% \end{figure}

We say that a source message consists of $K$ \emph{source packets} $\left\{\mathbf{s}_i\right\}_{i = 1}^K$ where $\mathbf{s}_i$ consists of $L$ elements of a finite field $\mathbb{F}_q$ of size $q$. A \emph{coded packet} $\mathbf{c}_j$ is also formed by $L$ elements from $\mathbb{F}_q$ and is defined as $\mathbf{c}_j = \sum_{i = 1}^K g_{i,j} \cdot \mathbf{s}_i$ where $g_{i,j} \in \mathbb{F}_q$ is referred to as a \emph{coding coefficient}. Provided that $N$ coded packets have been broadcast by the transmitter, the input to the broadcast channel can be expressed in matrix notation as $[ \mathbf{c}_1, \ldots, \mathbf{c}_N ] = [ \mathbf{s}_1, \ldots, \mathbf{s}_K ] \cdot \mathbf{G}$.
The $K \times N$ matrix $\mathbf{G}$ is defined by elements $g_{i,j}$, i.e., $\mathbf{G} \in \mathbb{F}_q^{K \times N}$. Coding coefficients are chosen at random over $\mathbb{F}_q$, in an identical and independent fashion according to the following probability law~\cite{7335581}:
\begin{equation}\label{eq.pl}
  \mathbb{P}\left(g_{i,j} = v\right) = \left\{ 
  \begin{array}{l l}
  	p & \quad \text{if $v = 0$}\\
    \displaystyle\frac{1-p}{q-1} & \quad \text{otherwise,}\\
  \end{array} \right.
\end{equation}
where $0 \leq p \leq 1$.  The greater the value of $p$, the more likely  that a coding coefficient is equal to $0$, so we observe that the average number of source packets actively participating in the generation of a coded packet is a function of $p$. The `classic' RLNC scheme refers to $p$ equal to $1/q$~\cite{7335581} (so the coding coefficients are uniform on $\mathbb{F}_q$), `sparse' RLNC schemes are characterized by $p > 1/q$.

Let $\left\{\mathbf{c}_j\right\}_{j = 1}^n$ be the set of coded packets that have been successfully received by a user, for $0 \leq n \leq N$. At the receiving end, each user populates a $K \times n$ decoding matrix $\mathbf{M}$ with the $n$ columns of $\mathbf{G}$ associated with the $n$ coded packets that have been successfully received.
Finally, relation $[ \mathbf{c}_1, \ldots, \mathbf{c}_n ] = [ \mathbf{s}_1, \ldots, \mathbf{s}_K ] \cdot \mathbf{M}$ holds.
The source message is recovered as soon as $\mathbf{M}$ becomes full rank and hence, $\mathbf{M}$ contains a $K \times K$ invertible matrix.

\section{Performance Analysis}\label{sec.PERF}
{Based on~\cite{6994250}, we observe that the probability of a user to recover a source message, i.e., the user delivery probability, as a function of $\epsilon$ can be expressed as follows:
\begin{equation} \label{eq:reps}
\mathrm{R}(\epsilon) = \sum_{n = K}^N \binom{N}{n} (1-\epsilon)^n \epsilon^{N-n} \mathrm{R}_{K,n}(p),
\end{equation}
where $\mathrm{R}_{K,n}(p)$ is the probability of a $K \times n$ decoding matrix being full rank,  as a function of $p$. In the case of classic RLNC, it is known that  $\mathrm{R}_{K,n}(p)_{\big|_{p = 1/q}} = \prod_{t = 0}^{K - 1} \left[1 - \frac{1}{q^{n-t}}\right]$ exactly~\cite{6994250}.
For sparse RLNC schemes, an exact expression for $\mathrm{R}_{K,n}(p)$ is still unknown but, as proposed in~\cite{7523985}, it can be approximated by means of the following lower-bound
\begin{equation}
\mathrm{R}_{K,n}(p) \geq 1\! -\! \min\left\{\!\eta_{\textrm{max}}(n); \sum_{t = 1}^{K} \!\binom{K}{t} (q-1)^{t-1} \rho_t \!\right\} \label{eq.sota.lb}
\end{equation}
and upper-bound
\begin{equation}
\mathrm{R}_{K,n}(p) \leq 1 \!-\! \max\left\{\!\eta_{\textrm{min}}(n); \sum_{t = 1}^{K}\! \binom{K}{t} p^{nt}(1-p^n)^{K-t} \!\right\} \label{eq.sota.ub}
\end{equation}
where
$\eta_{\textrm{max}}(t) = 1-\prod_{w = 0}^{K - 1} \left[1 - \left( \max \left\{p, \frac{1-p}{q-1}\right\} \right)^{t-w} \right]$ and $\eta_{\textrm{min}}(t) = 1-\prod_{w = 0}^{K - 1} \left[1 - \left( \min \left\{p, \frac{1-p}{q-1}\right\} \right)^{t-w} \right]$. From~\cite{RSA:RSA1}, it follows that $\eta_{\textrm{max}}(t)$ and $\eta_{\textrm{min}}(t)$ are the lower- and upper-bound to the probability, for $t \leq K$, of a $K \times t$ being non-full rank, respectively.
Finally, we write $\rho_\ell$ for the probability that any set of $\ell$ rows of a $K \times n$ matrix sums to the zero vector in $\mathbb{F}_q$, which can be expressed (directly following from~\cite[Eq.~(5)]{cooper1}), for $\ell = 1, \ldots, K$ and $n \geq K$, as:
\begin{equation} \label{eq:cooperprob}
\rho_\ell \doteq \left[ \frac{1}{q} \left( 1 + (q-1) \left( 1 - \frac{q(1-p)}{q-1} \right)^\ell  \right) \right]^n.
\end{equation}

Both~\eqref{eq.sota.lb} and~\eqref{eq.sota.ub} are based on $\eta_{\textrm{max}}(t)$ and $\eta_{\textrm{max}}(t)$, which essentially account for the event that some sets of rows  form a non-full rank matrix. Overall, $1-\eta_{\textrm{max}}(t)$ and $1-\eta_{\textrm{min}}(t)$ give a notoriously not tight approximation of $\mathrm{R}_{K,t}(p)$. This holds true especially for $K > 10$ and $p \geq 0.7$~\cite{7335581}. In addition, the right-hand terms in the minimization of~\eqref{eq.sota.lb} and in the maximization of~\eqref{eq.sota.ub} represent the probability of having any set of rows that linearly combined sums to the zero vector and the probability of having any groups of rows that are equal to the zero vector, respectively. In both cases, these events are significantly different to the event that some submatrix in $\mathbf{M}$  is not full rank -- thus impacting on the tightness of~\eqref{eq.sota.lb} and~\eqref{eq.sota.ub}. In the remainder of this section, we  address this issue by providing a novel expression for $\mathrm{R}_{K,n}(p)$, which tightly approximates the user delivery probability across a large range of system parameters.}

\vspace{-3mm}\subsection{Proposed Performance Model for Sparse RLNC}
Therefore, we consider the key research question: \emph{Given a $K \times n$ decoding matrix $\mathbf{M}$, formed according to the probability model \eqref{eq.pl}, what is the probability that $\mathbf{M}$ has rank $K$?} We remark that for $n < K$, the source message cannot be recovered, i.e., $\mathrm{R}_{K,n}(p)$ is equal to $0$. In the remainder of this section, we focus on the case where $n \geq K$  and we wish to know whether the $K$ rows of $\mathbf{M}$ form a linearly independent set, i.e., the rank of $\mathbf{M}$ is $K$.
We give the following definition.
\begin{definition} 
Write $\RRR \doteq \{ 1, 2, \ldots, \sum_{t=1}^K \binom{K}{t} \} = \{1, 2, \ldots,
 2^K - 1 \}$ for a set of labels. For each $r \in \RRR$, 
we regard $\set{S}_r$ as a subset of the set of indices $\{ 1, \ldots, K \}$ composed of $|\set{S}_r|$ items.
\end{definition}

It is immediate to prove that the following remark holds.
\begin{remark}\label{rem.ko}
Matrix $\mathbf{M}$ is full rank if and only if no linear combinations of any sets of rows indexed by a $\set{S}_r$ sums to the zero vector over the field $\mathbb{F}_q$. For $q = 2$, we can consider the
collection of events
\begin{equation} \label{eq:us}
U_{\set{S}_r} \doteq \left\{ \sum_{i \in \set{S}_r} \vc{m}_i = \vc{0} \right\}, \quad \text{for $r \in \RRR$},
\end{equation}
where we write $\vc{m}_i$ for the $i$-th row of $\mathbf{M}$, and where addition is understood to be over $\mathbb{F}_2$. We know that $\mathbf{M}$ is full rank if and only if none of the events $U_{\set{S}_r}$ occur for any $r \in \RRR$. \end{remark}

\begin{example} \label{ex:illustrate}
Consider the following matrix when $q=2$:
$$ \mathbf{M} = \left( \begin{array}{ccccccc}
1 & 0 & 0 & 1 & 1 & 0 & 1 \\
0 & 1 & 1 & 0 & 0 & 0 & 0 \\
1 & 0 & 1 & 0 & 0 & 1 & 1 \\
1 & 0 & 0 & 1 & 1 & 0 & 1 \\
1 & 1 & 0 & 0 & 0 & 1 & 1 
\end{array} \right). $$
In this case, rows 1 and 4 are identical, so $U_{\{1,4 \}}$ occurs. Further, rows 2,3 and 5 sum to zero over $\mathbb{F}_2$, so $U_{\{ 2,3,5 \}}$ also occurs. In addition, since both these sets of rows sum to zero, their union must also sum to zero, so $U_{ \{ 1,2,3,4,5 \}}$ also occurs.
\end{example}

{Example \ref{ex:illustrate} illustrates why it is not sufficiently accurate to estimate the full-rank probability of $\mathbf{M}$ by  considering the expected number of events $U_{\set{S}_r}$ which occur, using the expression for the probability of each individual $U_{\set{S}}$. This approach ignores the fact that such events are \emph{positively correlated}.  In general, given disjoint sets $\set{S}_1, \ldots, \set{S}_t$ such that $U_{\set{S}_1}, \ldots, U_{\set{S}_t}$ occur, then $U_{\set{S}}$ will occur for 
 each of the $2^{t}-1$ sets $\set{S}$ formed as unions of the $\set{S}_i$}

Our proposed performance framework builds upon a different set of statistical events, defined as follows.
\begin{definition}\label{def.h}
Let $V_{\set{S}_r}$ be defined as follows
\begin{equation} \label{eq:vs}
V_{\set{S}_r} \doteq U_{\set{S}_r} \bigcap \left( \bigcap_{\set{T} \subset \set{S}_r} U_{\set{T}}^C \right)\!\!, \,\,\,\,\text{for $r \in \RRR$},
\end{equation}
which is the event that the rows indexed by $\set{S}_r$ sum to the zero vector in $\mathbb{F}_2$ but that no collection of rows indexed by a proper subset of $\set{S}_r$ sums to the zero vector.
\end{definition}

In general, from Definition~\ref{def.h}, we have the following remark.

\begin{remark}\label{patch}
Matrix $\mathbf{M}$ is full rank if and only if none of the events $V_{\set{S}_r}$ occurs, for $r \in \RRR$.  
This choice of events significantly mitigates the impact of the  correlation among events observed in Example \ref{ex:illustrate}. There, $V_{\{1,4 \}}$ and $V_{\{ 2,3,5 \}}$ both occur (since no subset of them sums to zero), however $V_{ \{ 1,2,3,4,5 \}}$ does not.
This enables us to derive a tighter approximation of $\mathrm{R}_{K,n}(p)$.
\end{remark}

The proposed derivation of $\mathrm{R}_{K,n}(p)$ involves two approximation steps: (i) We approximate the probability  of event $V_{\set{S}_r}$ happening for any set $\set{S}_r$ consisting of a given number of items, and (ii) Since results based on the Stein-Chen method~\cite{arratia3,barbour} show the sum of approximately independent zero--one variables with small probability of being one is close to Poisson, we approximate $\mathrm{R}_{K,n}(p)$ with a negative exponential function. Firstly, we consider the following quantities:

\begin{definition}
For each $\ell = 1, \ldots, K$: 
\begin{enumerate}
\item For each $r \in \RRR$ such that the set $\set{S}_r$ has cardinality $\ell$,  the event $V_{\set{S}_r}$ has the same probability $\pi_\ell$ of happening, defined as
\begin{equation}
\pi_\ell \doteq \mathbb{P}\left[  V_{\set{S}_r}
\right], \quad \text{$\ell = 1, \ldots K$}.\label{eq.pi}
\end{equation}
\item We define a further quantity $\tilde{\pi}_\ell$ recursively as follows:
\begin{equation}
\tilde{\pi}_\ell \doteq \rho_\ell - \sum_{s=1}^{\ell-1} \binom{\ell-1}{s}  \rho_s \tilde{\pi}_{\ell-s}, \label{eq:recrelex}
\end{equation}
where (since taking $\ell = 1$ gives an empty sum) $\tilde{\pi}_1 \doteq \rho_1$.
\end{enumerate}
\end{definition}

\begin{lemma} \label{prop:vprob}
Term $\pi_\ell$ (defined in \eqref{eq.pi}) can be approximated as $\tilde{\pi}_\ell$ (defined in \eqref{eq:recrelex}).
\end{lemma}
\begin{IEEEproof}
See Appendix~\ref{app.A}.
\end{IEEEproof}

We observe that an obvious way in which $\mathbf{M}$ can fail to have full rank is that a particular row is identically zero. Indeed, considering this event gives an upper bound on $R_{K,n}(p)$, which for certain parameter values can be reasonably tight.
For this reason,  we condition out these events as follows:
\setlength{\arraycolsep}{0.0em}
\begin{align}
\mathrm{R}_{K,n}(p) & {}={} \pr(\{\mbox{$\mathbf{M}$ has no zero rows}\}) \notag\\
& \!\!\!\!\cdot \pr( \{\mbox{$\mathbf{M}$ is full rank}\} \;|\; \{\mbox{$\mathbf{M}$ has no zero rows}\}), \label{eq:zerorow}
\end{align}
where we can write the first term of \eqref{eq:zerorow} directly as $(1-p^n)^K$.

From Lemma~\ref{prop:vprob}, we prove the following result.
\begin{theorem} \label{th1}
We  approximate the second term of \eqref{eq:zerorow} as
\begin{equation}
\pr(\mbox{$\mathbf{M}$ is full rank} \;|\; \mbox{$\mathbf{M}$ has no zero rows}) \simeq \exp(-\lambda), \label{eq:thX}
\end{equation}
where
\begin{equation}
\lambda \doteq \sum_{\ell=2}^K \lambda_\ell, \quad\quad\text{for $\lambda_\ell \doteq \binom{K}{\ell} \frac{\tilde{\pi}_\ell}{(1 - p^n)^\ell}$,}\label{eq.lambL} 
\end{equation}
so we approximate $\mathrm{R}_{K,n}(p)$ as follows:
\begin{equation}
\mathrm{R}_{K,n}(p) \cong (1 - p^n)^K \exp(-\lambda). \label{eq.th2}
\end{equation}
\end{theorem}
\begin{IEEEproof}
See Appendix~\ref{app.A}.
\end{IEEEproof}

{The most computationally intensive part of calculating~\eqref{eq.th2} is the derivation of $\tilde{\pi}_\ell$, which requires $O(K^2)$ operations.
However, since the expression of $\tilde{\pi}_\ell$ is independent of $K$, it has to be computed only once to approximate $\mathrm{R}_{K,n}(p)$.
In addition, for $K = 10, 20, 50$ and $100$, the average time needed to compute\footnote{Tests performed by running our benchmark code on one core of an Intel Xeon CPU E5-2650v4 operated at \SI{2.20}{\GHz}.} $\tilde{\pi}_1, \ldots, \tilde{\pi}_K$ (normalized by $K$) is equal to $2.7 \cdot 10^{-3}\SI{}{\second}$, $7.7 \cdot 10^{-3}\SI{}{\second}$, $7.3 \cdot 10^{-2}\SI{}{\second}$ and $4.3 \cdot 10^{-1}\SI{}{\second}$, respectively.
It is also key to note that Theorem~\ref{th1} allows us to decouple the impact that any $\ell \times n$ submatrix of $\mathbf{M}$ has on the approximation of $\mathrm{R}_{K,n}(p)$ as in~\eqref{eq.th2}, for $\ell = 2, \ldots, K$. As  the following remark explains, this  allows us to further approximate~\eqref{eq.th2} by reducing the number of summation terms defining $\lambda$ and hence, reducing the computational complexity of the approximation~\eqref{eq.th2}.}
\begin{remark}\label{rem}
Consider the set of all the $\ell \times n$ submatrices of $\mathbf{M}$, then $\lambda_\ell$ approximates the probability that at least one of these submatrices is not full rank, assuming $\mathbf{M}$ has no zero rows.
For this reason, the approximation of $\mathrm{R}_{K,n}(p)$ given in~\eqref{eq.th2} can be further approximated by referring to those submatrices of $\mathbf{M}$ composed by up to $m$ rows, for $m = 2, \ldots, K$. As such, with define the $m$-th approximation order of~\eqref{eq.th2} as follows:
\begin{equation}
\mathrm{R}_{K,n}^{(m)}(p) \doteq (1 - p^n)^K \exp\left(-\sum_{\ell = 2}^m \lambda_\ell\right).\label{eq.orderRem}
\end{equation}
\end{remark}

Let us consider the following approximation order optimization (AOO) problem\footnote{For the sake of compactness, with a slight abuse of notation, we say that $\mathrm{R}_{K,n}^{(m)}(p)$ is always equal to $\mathrm{R}_{K,n}^{(K)}(p)$, for any $m > K$.}:
\begin{align}
	 \text{AOO} &  \quad  \min_{m \in \{2, \ldots, K\} } \,\,  m \label{OP.of}\\
    \text{s.t.} &   \quad e(m) \leq \tau \;\bigvee\; m \leq \Hat{m}\label{OP.c1}
\end{align}
where function $e(m)$ is defined as $\mathrm{R}_{K,n}^{(m)}(p) - \mathrm{R}_{K,n}^{(m+1)}(p)$. The solution $m^*$ to the AOO problem represents the smallest-order approximation of~\eqref{eq.th2} associated with a target error value $\tau \in [0, 1]$ or such that $m^*$ is smaller than $\Hat{m}$, for $2 \leq \Hat{m} \leq K$.
\begin{remark}\label{rem.1}
From~\eqref{eq.lambL}, it follows that term $\sum_{\ell = 2}^m \lambda_\ell$ is a non-decreasing function of $m$, i.e., relation $\mathrm{R}_{K,n}^{(m)}(p) \geq \mathrm{R}_{K,n}^{(m+1)}(p) \geq \mathrm{R}_{K,n}^{(K)}(p)$ holds. As such, for any given $K$, $n$ and $p$, the error function $e(m)$ attains only one maximum, for $m \in \{2, \ldots, K\}$. For this reason, the AOO problem can be solved iteratively evaluating $\mathrm{R}_{K,n}^{(m^*)}(p)$, for $m^* = 2, \ldots, K$, until $e(m^*) \geq e(m^* + 1)$ and $e(m^* + 1) \leq \tau$ or $m^*$ is smaller than or equal to $\hat{m}$.
\end{remark}

\begin{remark}\label{AnyQ}
From Remark~\ref{patch}, $\mathbf{M}$ is full rank if and only if none of the events $V_{\set{S}_r}$ occurs, for $r \in \RRR$, and $q = 2$. However, for non-binary fields, the aforementioned statement captures a subsets of events when a random matrix is full rank. As such, we propose to use~\eqref{eq.th2} to approximate $\mathrm{R}_{K,n}(p)$, for $q > 2$.
\end{remark}

\vspace{-3mm}\section{Analytical Results}\label{sec.RES}
This section compares the approximation we proposed in Theorem~\ref{th1} against the approximation~\eqref{eq.sota.lb} and~\eqref{eq.sota.ub}. Both our simulator and the implementation of the proposed theoretical framework are available online~\cite{SimFW}.

\begin{figure}[tb]
\centering
	\includegraphics[width=0.95\columnwidth]{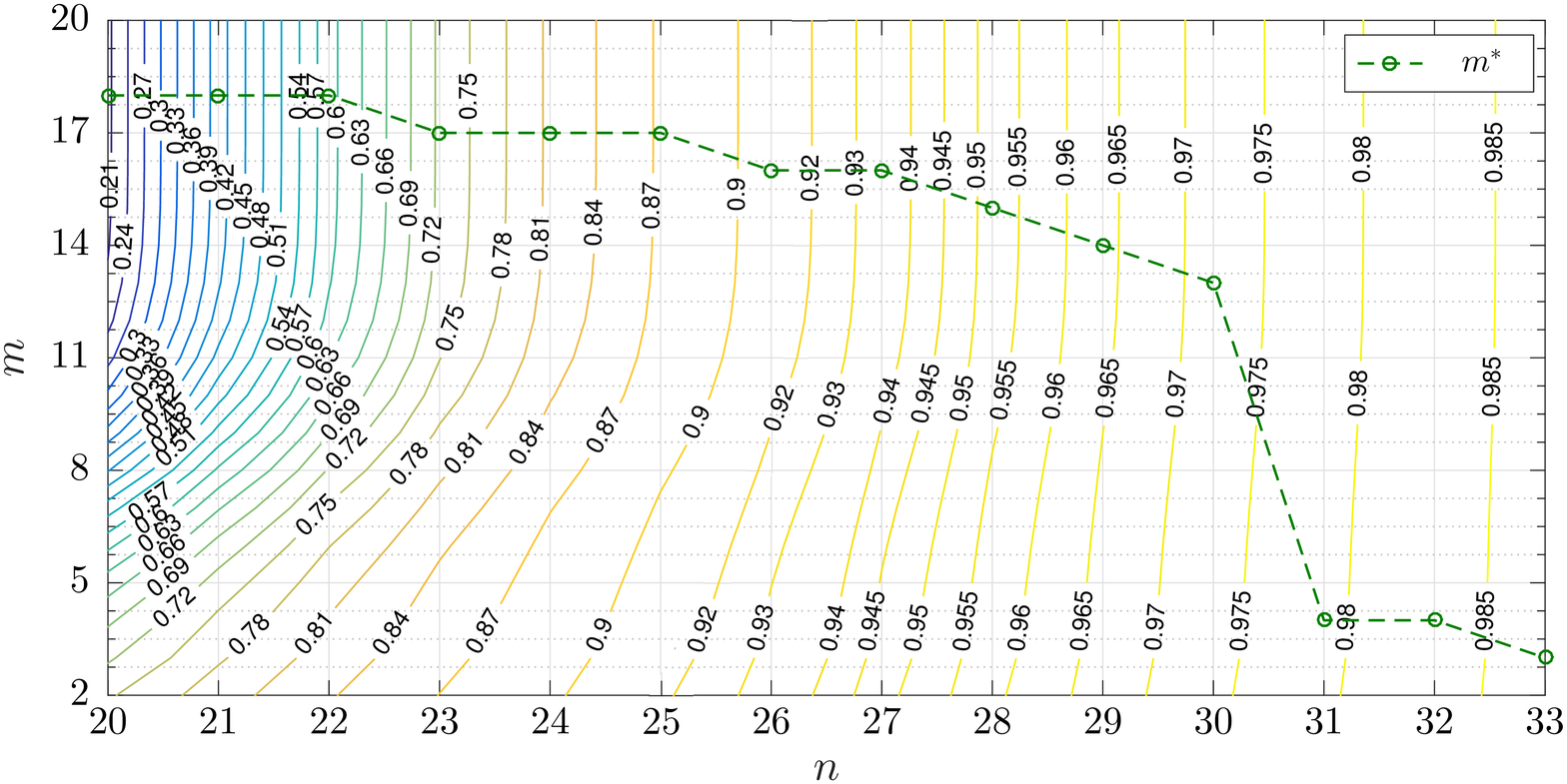}\vspace{-3mm}
\caption{Contour map of $\mathrm{R}_{K,n}^{(m)}$ as a function of $n$ and $m$, for $q = 2$, $K = 20$ and $p = 0.8$. The values of $m^*$ have been derived by referring to $\tau = 10^{-4}$ and $\hat{m} = K$.}\vspace{-3mm}
\label{fig.f_1}
\end{figure}

Fig.~\ref{fig.f_1} shows the relationship that exists between the order $m$ of the approximation as in~\eqref{eq.orderRem} and the number $n$ of received coded packets in $\mathrm{R}_{K,n}^{(m)}$, for $q = 2$, a source message composed by $K = 20$ packets and $p = 0.8$. From~\eqref{eq.orderRem}, we remark that, for a given value of $n$, $\mathrm{R}_{K,n}^{(m)}$ is a non-increasing function of $m$. This is directly related to the fact that small approximation orders account for submatrices of $\mathbf{M}$ composed of a reduced number of rows. This can be intuitively explained by considering the extreme case where $n$ is large compared to $K$. In this case, the probability of $\mathbf{M}$ being full rank can be approximated by the probability of having  a set of $K$ (non-zero) rows of $\mathbf{M}$ where no rows are identical -- this corresponds to the case where $m$ is set equal to $2$.

The aforementioned facts are confirmed by Fig.~\ref{fig.f_1}. For instance, for $n = 20$, the value of $\mathrm{R}_{K,n}^{(m)}$ drops from $0.72$ ($m = 2$) to $0.21$ ($m = 14$) to remain almost unchanged for $14 \leq m \leq 20$. In particular, by solving the AOO problem for $\tau = 10^{-4}$, $\hat{m} = K$ and $n = 20$, we obtain an optimal value of $m^*$ equal to $18$ as per Remark~\ref{rem}.
We also observe that the value of $m^*$ appears to sharply decrease as $n$ increases, which makes computationally convenient to approximate $\mathrm{R}_{K,n}$ with $\mathrm{R}_{K,n}^{(m^*)}$. For instance, Fig.~\ref{fig.f_1} shows that the error function $e(m^*)$ takes values smaller than or equal to $\tau = 10^{-4}$ for $n = 31$ and $m^* = 4$ -- thus making it pointless to approximate $\mathrm{R}_{K,n}$ with  an heuristic order equal to or greater than $5$. In the remainder of this section, to highlight the accuracy of our approximation, we will refer to a value of $\tau = 10^{-10}$ or $\hat{m} \leq \lceil 3K/4 \rceil$.

\begin{figure}[tb]
\centering
\subfloat[$p = 0.7$]{\label{fig.f_2.1}
	\includegraphics[width=\columnwidth]{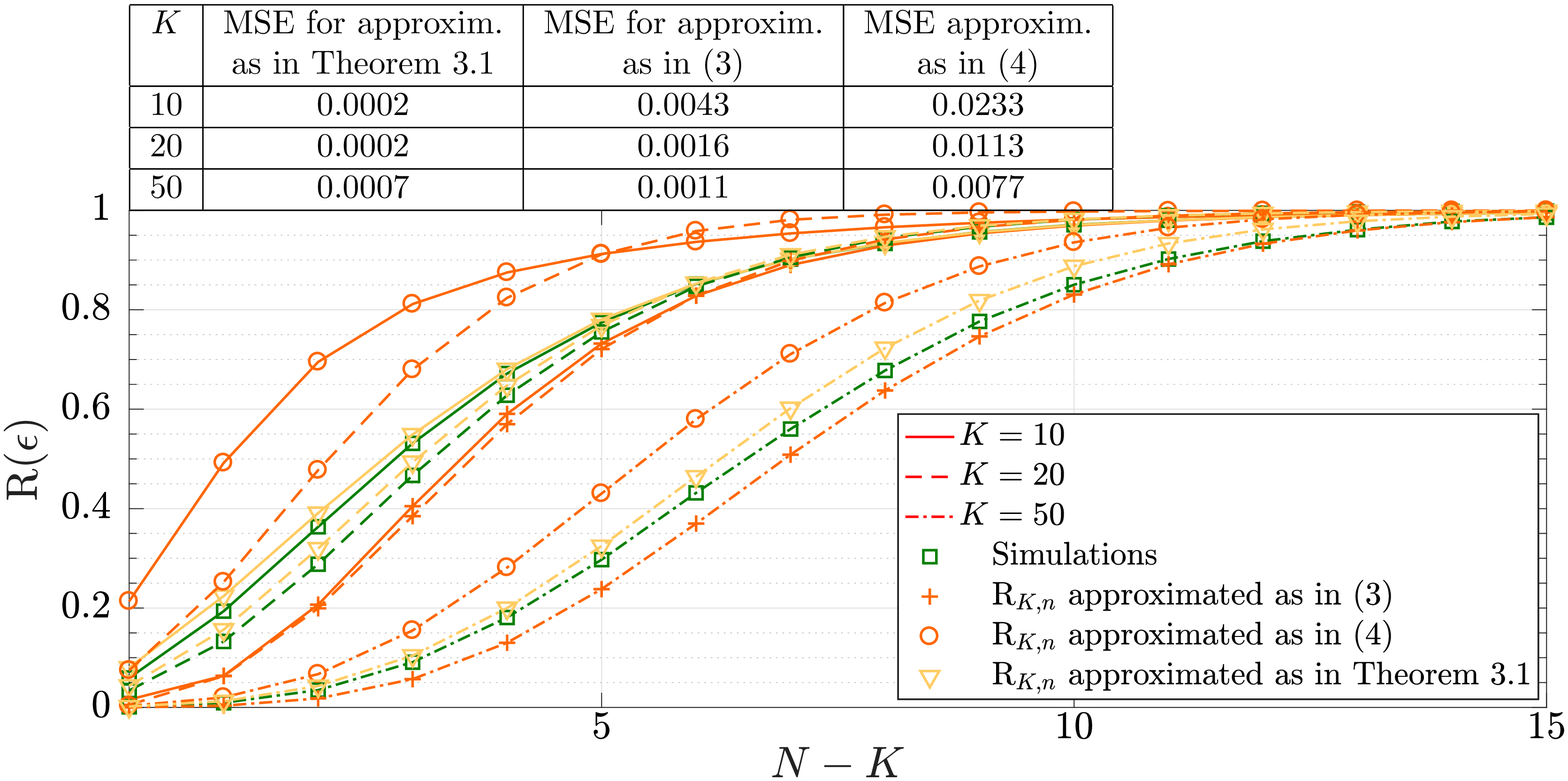}
}\vspace{-3mm}\\
\subfloat[$p = 0.9$]{\label{fig.f_2.2}
	\includegraphics[width=\columnwidth]{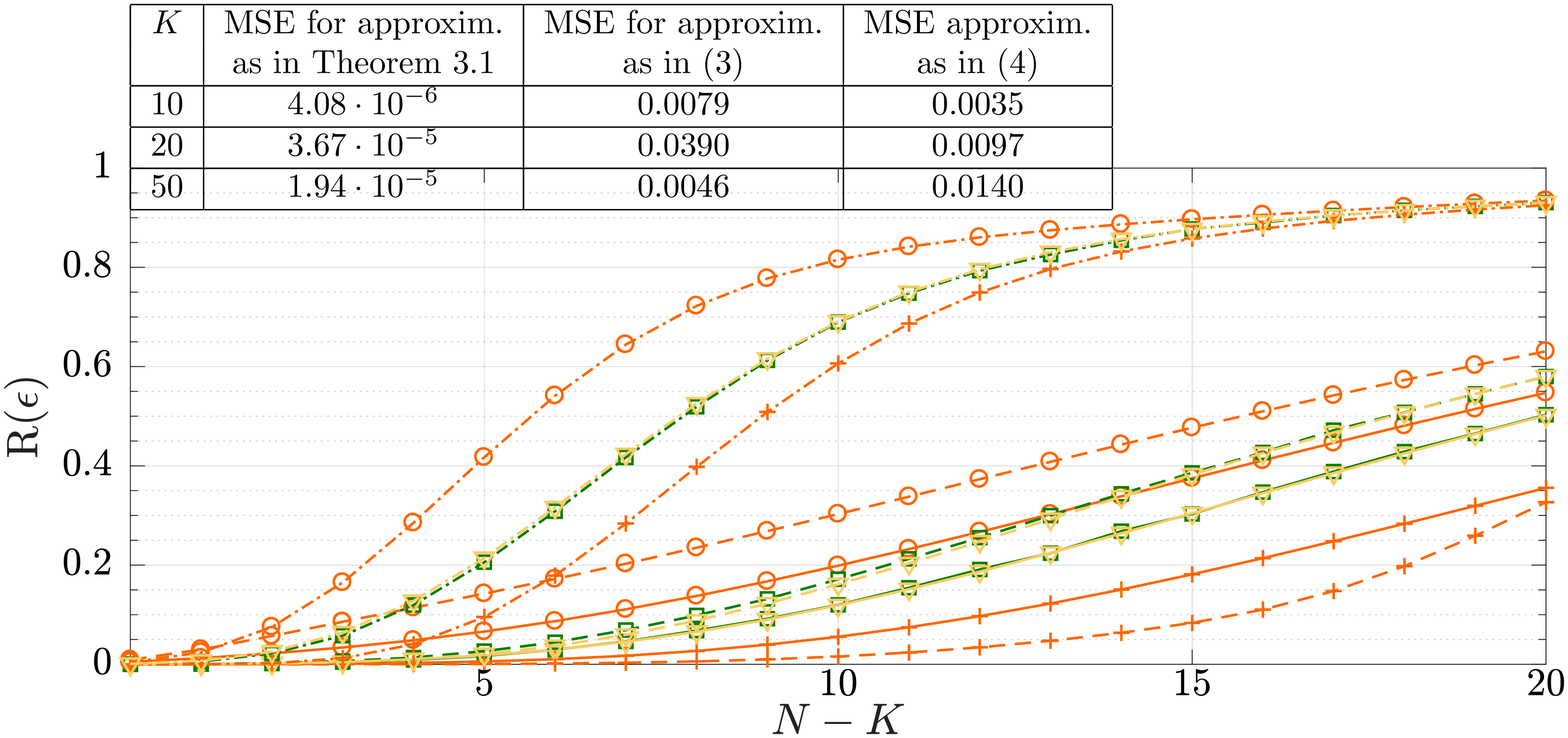}
}
\vspace{-3mm}\caption{Probability $\mathrm{R}(\epsilon)$ of recovering a source message of $K = \{10,20,50\}$ source packets as a function of the number of coded packet transmissions $N$, for $q = 2$ and $\epsilon = 0.1$. Tables shows the MSEs between simulation results and the cases where $\mathrm{R}_{K,n}$ is approximated as in~\eqref{eq.sota.lb},~\eqref{eq.sota.ub} and Theorem~\ref{th1}. Legend of both figures is reported in Fig.~\ref{fig.f_2.1}.}\vspace{-5mm}
\label{fig.f_2}
\end{figure}

\begin{figure}[tb]
\vspace{-9mm}\centering
\subfloat[$p = 0.7$]{\label{fig.f_5.1}
	\includegraphics[width=0.39\columnwidth]{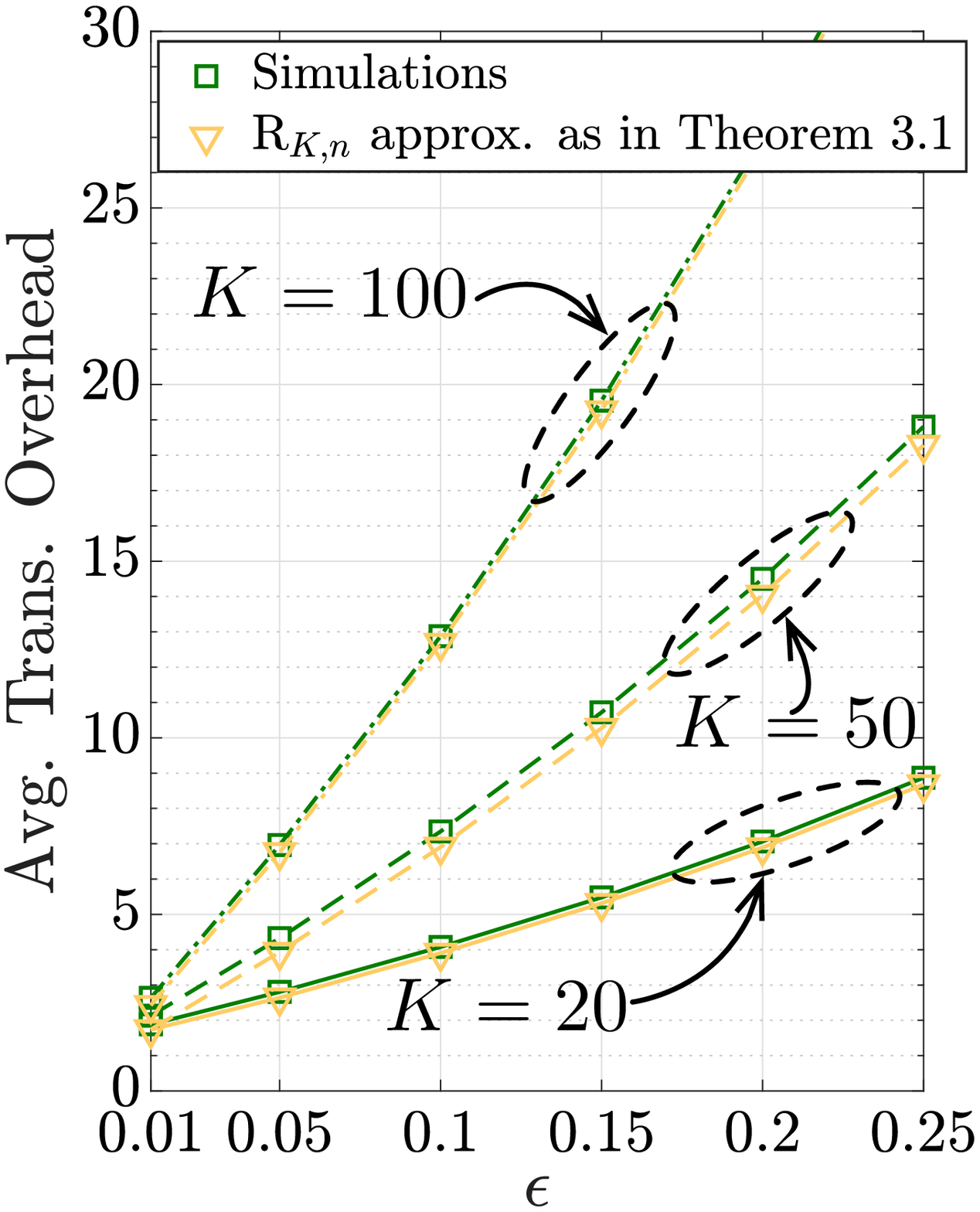}
}
\subfloat[$p = 0.9$]{\label{fig.f_5.2}
	\includegraphics[width=0.39\columnwidth]{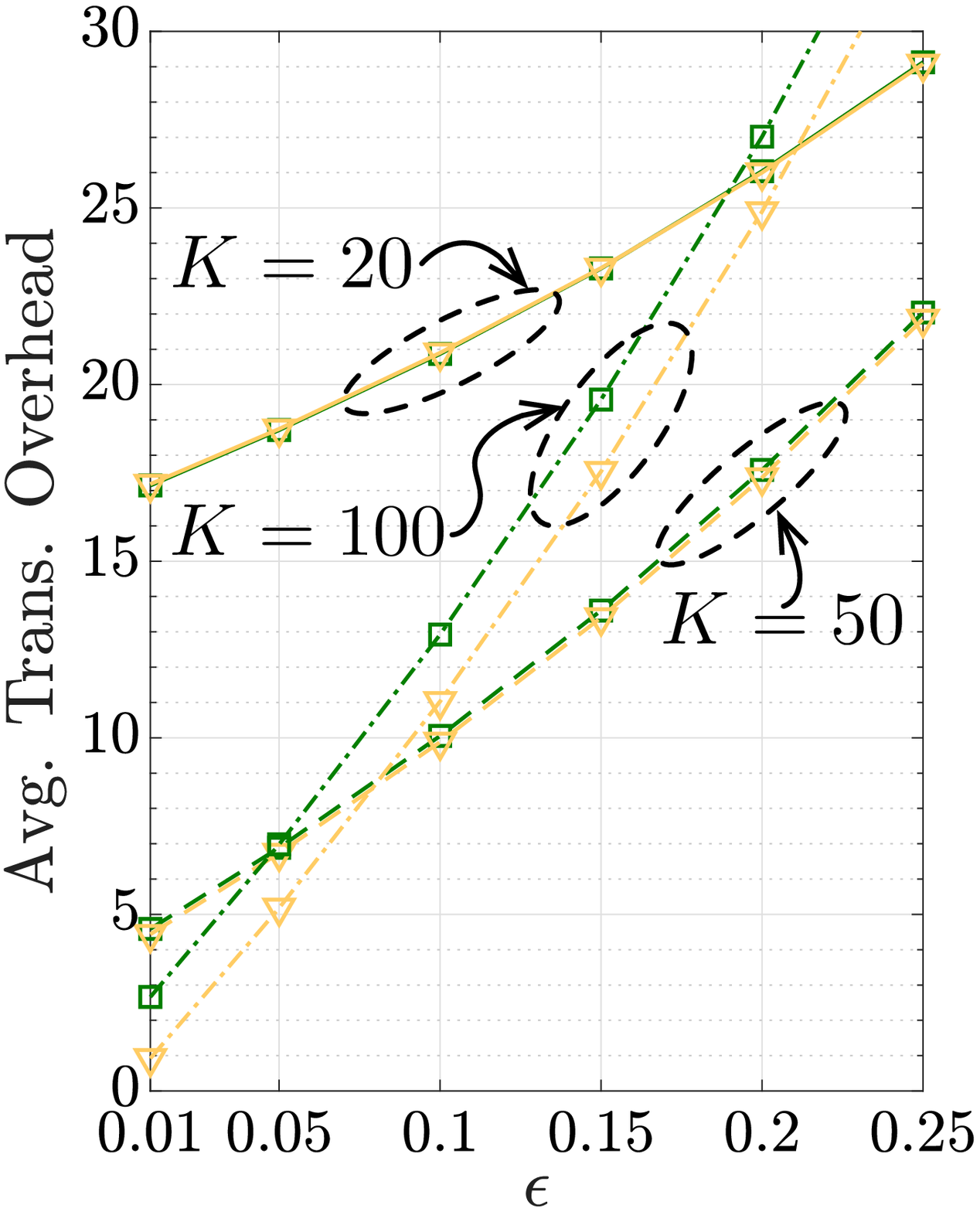}
}\vspace{-1mm}
\caption{Average transmission overhead as a function of $\epsilon$, for \mbox{$K = \{20,50,100\}$}, $q = 2$ and $p = \{0.7,0.9\}$. Legend of both figures is reported in Fig.~\ref{fig.f_5.1}.}\vspace{-5mm}
\label{fig.f_5}
\end{figure}

{Fig.~\ref{fig.f_2} compares the user delivery probability $\mathrm{R}(\epsilon)$ for $K = 10, 20$ and $50$ and $q=2$. We compare the value for $\mathrm{R}(\epsilon)$ implied by \eqref{eq:reps}, substituting the approximations to $\mathrm{R}_{K,n}$ given by~\eqref{eq.sota.lb},~\eqref{eq.sota.ub} and our proposal in~\eqref{eq.orderRem} to the probability $\mathrm{R}(\epsilon)$ estimated by  Monte Carlo simulations. Results are given as a function of $N-K$, which represents the transmission overhead, i.e., the number of coded packets in excess of $K$ that are transmitted. Assuming that the time needed to transmit each coded packet is fixed and equal to one time slot, the goodput of the system can be immediately expressed as the bit length of the source message divided by the time duration of $N$ time slots.
For concreteness, we considered a value of packet error probability $\epsilon = 0.1$, which is the maximum transport block error probability regarded as acceptable in a Long Term Evolution-Advanced (LTE-A) system~\cite{6994250}. In particular, in the case where $p = 0.7$, Fig.~\ref{fig.f_2.1} shows that the maximum gap between our proposed approximation~\eqref{eq.orderRem} and simulation results is equal to $3.1 \cdot 10^{-2}$, which occurs for $K = 20$. Fig.~\ref{fig.f_2.2} refers to the case when $p = 0.9$ and shows that the gap between ~\eqref{eq.orderRem} and simulation results is negligible.

In contrast, both Fig.~\ref{fig.f_2.1} and~\ref{fig.f_2.2} show that approximating $\mathrm{R}_{K,n}$ using the state of the art~\eqref{eq.sota.lb} and~\eqref{eq.sota.ub} leads $\mathrm{R}(\epsilon)$ to significantly deviate from the simulation results. For instance, for $p = 0.9$, $K = 50$ and $N-K = 7$, the absolute deviation can be up to $0.14$ and $0.22$, in the case of~\eqref{eq.sota.lb} and~\eqref{eq.sota.ub}, respectively.
In general, the maximum Mean Squared Error (MSE) between simulations and our proposed approximation~\eqref{eq.orderRem} is experienced for $K = 50$ and $p = 0.7$ and it is equal to $7 \cdot 10^{-4}$. That is smaller than the corresponding MSEs between simulation and approximations~\eqref{eq.sota.lb} and~\eqref{eq.sota.ub}, which are equal to $1.1 \cdot 10^{-3}$ and $7.7 \cdot 10^{-3}$, respectively (between $1.6$ and $11$ times smaller). In addition, Fig.~\ref{fig.f_2.2} shows that, for $p = 0.9$, our proposal overlaps simulation results while the considered alternatives significantly deviate. In this case, the MSE of our approximation is between $237$ times (for $K = 50$) and $1063$ times ($K = 20$) smaller than in the case of~\eqref{eq.sota.lb} and, between $722$ times ($K = 50$) and $857$ times (for $K = 10$) smaller than in the case of~\eqref{eq.sota.ub}.

\begin{figure}[t]
\centering
\includegraphics[width=\columnwidth]{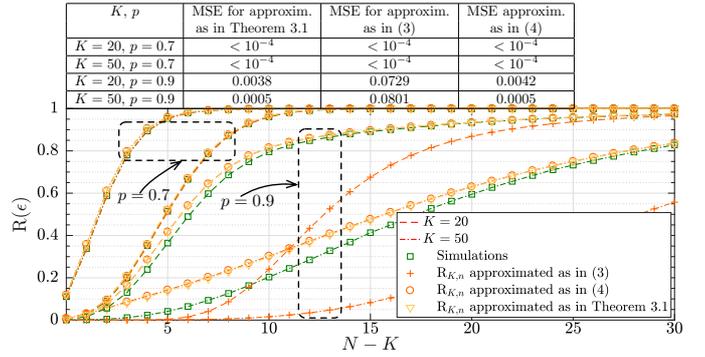}
\vspace{-3mm}\caption{Probability $\mathrm{R}(\epsilon)$ of recovering a source message of $K = \{20,50\}$ source packets as a function of the number of coded packet transmissions $N$, for $p = \{0.7, 0.9\}$, $q = 2^4$ and $\epsilon = 0.1$. Tables shows the MSE between simulation results and the cases where $\mathrm{R}_{K,n}$ is approximated as in~\eqref{eq.sota.lb},~\eqref{eq.sota.ub} and Theorem~\ref{th1}.}\vspace{-4mm}
\label{fig.f_3}
\end{figure}

Fig.~\ref{fig.f_3} shows that for $q = 2^4$, our approximation either overlaps (for $p = 0.7$) or marginally diverges from simulation results ($p = 0.9$). In the latter case, we observe that the (absolute gap) never exceeds $0.11$ and the maximum MSE is equal to $3.8 \cdot 10^{-3}$.
Similar behavior is also exhibited by the case where $\mathrm{R}_{K,n}$ is approximated as in~\eqref{eq.sota.lb} or~\eqref{eq.sota.ub} and $p = 0.7$. However, as $p$ increases to $0.9$, the approximation based on~\eqref{eq.sota.lb} significantly deviates from simulation results even of a quantity larger than $0.51$ ($K = 20$ and $N-K = 9$).

Generally, from Figs.~\ref{fig.f_2} and~\ref{fig.f_3} we observe that, for binary fields (with the only exception of $K = 50$ and $p = 0.9$), the approximation based on~\eqref{eq.sota.lb} is tighter than that based on~\eqref{eq.sota.ub}. However, the exact opposite holds as both $q$ and $p$ increase. Our proposed approximation avoids this issue. In fact, our solution tightly approximates simulation results, for all the cases considered. These conclusions are also confirmed by Fig.~\ref{fig.f_4}, which shows the probability $\mathrm{R}(\epsilon)$ as a function of $p$, for $q = \{2,2^4\}$, $K = \{20, 50\}$.

As an immediate application of a tighter approximation of the user delivery probability, we can accurately estimate the average transmission overhead needed for a user to recover a source message (that is $\sum_{t = K}^\infty t \cdot \mathrm{R}_{|_{N = t}}(\epsilon) - K$). In particular, Fig.~\ref{fig.f_5} shows the average transmission overhead as a function of $\epsilon$, for $K = 20, 50$ and $100$ and, $p = 0.7$ and $0.9$. Direct proof of the quality of the proposed approximation is given by the fact that the deviation between theoretical and simulation results is negligible across the whole range of parameters -- the maximum gap between theory and simulations is equal to $2$ and occurs for $K = 100$ and $p = 0.9$.}

\section{Conclusion}\label{sec.CL}{
This paper presented a novel approximated performance model for a sparse RLNC implementation. The proposed model exploits the Stein-Chen method to derive a tight approximation to the probability of a user recovering a source message.
Analytical results show that the Mean Squared Error (MSE) between our approximation and simulation results, for $q = 2$ and $2^4$, never exceeds $7 \cdot 10^{-4}$ and $3.8 \cdot 10^{-3}$, respectively. On the other hand, the state of the art bounds are not always tight. For instance, when $q = 2$ our proposal is between $1.5$ and $1063$ times closer in MSE to simulations.
}

\vspace{-2mm}\appendices\section{}\label{app.A}{
\vspace{-2mm}\begin{IEEEproof}[Proof of Lemma~\ref{prop:vprob}]
By symmetry it is enough to consider a subset of rows of the form $\set{S} = \{ 1, 2, \ldots, \ell \}$, where $\ell \leq K$. The key is to fix one row (say row $1$) and to consider the smallest set of rows containing row $1$ which sums to zero. Consider $\set{W}$, a subset of $\set{S}$ with $1 \in \set{W}$, and say that event $T_{\set{W},\set{S}}$ occurs when both: 
(i) the rows of $\mathbf{M}$ with indices in $\set{S}$ add to zero and,
(ii)  rows with indices in $\set{W}$ add to zero, but no subset of these rows add to zero, i.e. event $V_{\set{W}}$ occurs see \eqref{eq:vs}.
By considering (i) and (ii) together, $T_{\set{W}, \set{S}}$ occurs when the rows in both the sets  $\set{S} \setminus \set{W}$ and $\set{W}$ (but no subset of $\set{W}$) add to zero. In other words $T_{\set{W},\set{S}}$  equals $U_{\set{S} \setminus \set{W}} \bigcap V_{\set{W}} $. Since rows in $\mathbf{M}$ are statistically independent, for each $\set{W}$ of size $(\ell-s)$, the event $T_{\set{W},\set{S}}$ occurs with probability 
\vspace*{-2mm}\begin{align}
\pr( T_{\set{W},\set{S}}) &= \pr \left( U_{\set{S} \setminus \set{W}} \bigcap V_{\set{W}} \right)\notag\\
&= \pr \left( U_{\set{S} \setminus \set{W}} \right) \pr \left( V_{\set{W}} \right) = \rho_s \pi_{\ell-s}.\vspace*{-1mm}
\end{align}
Furthermore,  since $U_{\set{S}} = \bigcup_{\set{W}} T_{\set{W},\set{S}}$ for any $\set{S}$
we observe:
\begin{align}
\vspace*{-2mm}\rho_l & = \pr( U_{\set{S}} )  =  \pr \left( \bigcup_{\set{W}} T_{\set{W},\set{S}} \right) 
\simeq \sum_{\set{W}} \pr( T_{\set{W},\set{S}} ) \notag\\
& = \sum_{s=0}^{\ell-1} \binom{\ell-1}{s}  \rho_s \tilde{\pi}_{\ell-s}  
 = \sum_{s=1}^{\ell-1} \binom{\ell-1}{s}  \rho_s \pi_{\ell-s} + \pi_\ell.\vspace*{-2mm}
\end{align}
This relation holds because (i) we assume events $T_{\set{W},\set{S}}$ are approximately disjoint (ii)
there are $\binom{\ell-1}{s}$ possible sets $\set{W}$ of size $(\ell-s)$ containing 1. In Example \ref{ex:illustrate}, if $\set{S} = \{ 1,2,3,4,5 \}$ then $U_{\set{S}}$ occurs
as previously discussed, indeed so does $T_{\set{W},\set{S}}$ with $\set{W} = \{1,4 \}$.
This concludes the proof. \end{IEEEproof}

\begin{figure}[t]
\centering
	\vspace{-4mm}\includegraphics[width=\columnwidth]{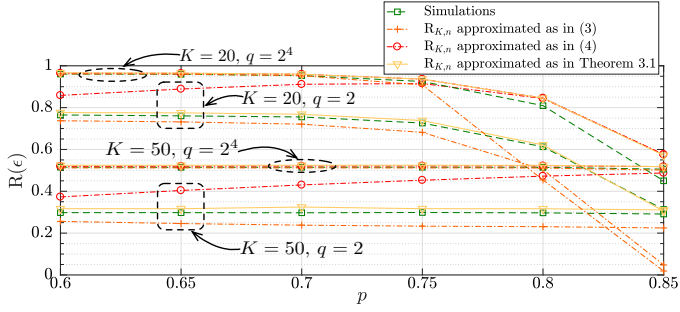}\vspace{-3mm}
\caption{Probability $\mathrm{R}(\epsilon)$ of recovering a source message of $K = 20$ and $50$ source packets for $N = 25$ and $55$, respectively, as a function of $p$ with $q = \{2,2^4\}$ and $\epsilon = 0.1$.}\vspace{-5mm}
\label{fig.f_4}
\end{figure}

\begin{IEEEproof}[Proof of Theorem~\ref{th1}]
We write $\RRR^*$ for the collection of indices $r$ such that $| \set{S}_r |  \geq 2$ and define the random variable
\begin{equation}
\vspace*{-2mm}W \doteq \sum_{r \in \RRR^*} \ind{V_{\set{S}_r}\;|\;\mbox{$\mathbf{M}$ has no zero rows}} \vspace*{0mm}
\end{equation}
where indicator function $\ind{\cdot}$ equals  $1$ if a particular event has occurred, or $0$ otherwise.  Observe that
$ \pr(\mbox{$\mathbf{M}$ is full rank}\;|\;\{\mbox{$\mathbf{M}$ has no zero rows}\}) = \pr(W = 0)$.
By~\eqref{eq.pi},  if $\set{S}_r$ has $\ell \geq 2$ elements, independence of the rows means that the probability 
\vspace*{-0.5mm}\begin{align}
& \pr  \left( V_{\set{S}_r}\;|\;\{\mbox{no zero rows in $\mathbf{M}$}\}\right) \notag\\
& =  \frac{ \pr \left( V_{\set{S}_r} \bigcap \{ \mbox{no zero rows in $\mathbf{M}$} \}  \right)}{(1-p^n)^K} \notag\\
& = \frac{ \pr \left( V_{\set{S}_r} \right) \pr \left( \{ \mbox{no zero rows in $\set{S}_r^c$} \} \right)}{(1-p^n)^K} 	\notag\\
& \cong \frac{\tilde{\pi}_\ell (1-p^n )^{K-\ell}}{(1-p^n)^K} 
= \frac{\tilde{\pi}_\ell}{(1 - p^n)^\ell}.
\end{align}
Hence, by counting sets of different sizes in $\RRR^*$, we it follows that $\ep[W] = \lambda$.
Further, $W$ is the sum of a large number of zero--one variables, each of which equals one with small probability, and where each random variable in the sum is independent of a large proportion of the other terms. These are the conditions under which $W$ is close to Poisson, as shown by the Stein--Chen method~\cite{arratia3,barbour}. Approximation~\eqref{eq.th2} follows because~\cite[Theorem 1]{arratia3} means that $\pr(W = 0) \simeq \exp(-\lambda)$.
\end{IEEEproof}}

\bibliographystyle{IEEEtran}
\vspace{-6mm}\bibliography{IEEEabrv,papers}

\begin{thebibliography}{10}
\providecommand{\url}[1]{#1}
\csname url@samestyle\endcsname
\providecommand{\newblock}{\relax}
\providecommand{\bibinfo}[2]{#2}
\providecommand{\BIBentrySTDinterwordspacing}{\spaceskip=0pt\relax}
\providecommand{\BIBentryALTinterwordstretchfactor}{4}
\providecommand{\BIBentryALTinterwordspacing}{\spaceskip=\fontdimen2\font plus
\BIBentryALTinterwordstretchfactor\fontdimen3\font minus
  \fontdimen4\font\relax}
\providecommand{\BIBforeignlanguage}[2]{{%
\expandafter\ifx\csname l@#1\endcsname\relax
\typeout{** WARNING: IEEEtran.bst: No hyphenation pattern has been}%
\typeout{** loaded for the language `#1'. Using the pattern for}%
\typeout{** the default language instead.}%
\else
\language=\csname l@#1\endcsname
\fi
#2}}
\providecommand{\BIBdecl}{\relax}
\BIBdecl

\bibitem{5G-PPP}
\BIBentryALTinterwordspacing
``{5G-PPP White Paper on Automotive Vertical Sector},'' 5G Infrastructure
  Public Private Partnership, Tech. Rep., Oct. 2015. [Online]. Available:
  \url{https://5g-ppp.eu/wp-content/uploads/2014/02/5G-PPP-White-Paper-on-Automotive-Vertical-Sectors.pdf}
\BIBentrySTDinterwordspacing

\bibitem{6416071}
E.~Magli, M.~Wang, P.~Frossard, and A.~Markopoulou, ``{Network Coding Meets
  Multimedia: A Review},'' \emph{{IEEE} Trans. Multimedia}, vol.~15, no.~5, pp.
  1195--1212, 2013.

\bibitem{7393818}
P.~Wang, G.~Mao, Z.~Lin, M.~Ding, W.~Liang, X.~Ge, and Z.~Lin, ``{Performance
  Analysis of Raptor Codes Under Maximum Likelihood Decoding},'' \emph{{IEEE}
  Trans. Commun.}, vol.~64, no.~3, pp. 906--917, Mar. 2016.

\bibitem{6994250}
A.~Tassi, I.~Chatzigeorgiou, and D.~Vukobratovi\'c, ``{Resource-Allocation
  Frameworks for Network-Coded Layered Multimedia Multicast Services},''
  \emph{{IEEE} J. Sel. Areas Commun.}, vol.~33, no.~2, pp. 141--155, Feb. 2015.

\bibitem{EvTassi}
\BIBentryALTinterwordspacing
E.~Tsimbalo, A.~Tassi, and R.~J. Piechocki, ``{Reliability of Multicast under
  Random Linear Network Coding},'' Oct. 2017. [Online]. Available:
  \url{http://arxiv.org/abs/1709.05477}
\BIBentrySTDinterwordspacing

\bibitem{DanielTCOM}
P.~Garrido, D.~E. Lucani, and R.~Ag\"{u}ero, ``{Markov Chain Model for the
  Decoding Probability of Sparse Network Coding},'' \emph{{IEEE} Trans.
  Commun.}, vol.~65, no.~4, pp. 1675--1685, Apr. 2017.

\bibitem{ComplexityGE}
D.~Andr\'en, L.~Hellstr{\o}m, and K.~Markstr{\o}m, ``{On the Complexity of
  Matrix Reduction Over Finite Fields},'' \emph{Advances in Applied
  Mathematics}, vol.~39, no.~4, pp. 428--452, 2007.

\bibitem{7335581}
A.~Tassi, I.~Chatzigeorgiou, and D.~E. Lucani, ``{Analysis and Optimization of
  Sparse Random Linear Network Coding for Reliable Multicast Services},''
  \emph{{IEEE} Trans. Commun.}, vol.~64, no.~1, pp. 285--299, Jan. 2016.

\bibitem{cooper1}
C.~Cooper, ``On the distribution of rank of a random matrix over a finite
  field,'' \emph{Random Struct. Alg.}, vol.~17, no. 3-4, pp. 197--212, 2000.

\bibitem{RSA:RSA1}
J.~Bl\"{o}mer, R.~Karp, and E.~Welzl, ``{The Rank of Sparse Random Matrices
  Over Finite Fields},'' \emph{Random Struct. Alg.}, vol.~10, no.~4, pp.
  407--419, 1997.

\bibitem{7523985}
A.~S. Khan and I.~Chatzigeorgiou, ``{Improved Bounds on the Decoding Failure
  Probability of Network Coding Over Multi-Source Multi-Relay Networks},''
  \emph{{IEEE Commun. Lett.}}, vol.~20, no.~10, pp. 2035--2038, Oct. 2016.

\bibitem{5978939}
X.~Li, W.~H. Mow, and F.-L. Tsang, ``{Rank Distribution Analysis for Sparse
  Random Linear Network Coding},'' in \emph{Proc. of NetCod 2011}, Beijing,
  China, CN, Jul. 2011, pp. 1--6.

\bibitem{arratia3}
R.~Arratia, L.~Goldstein, and L.~Gordon, ``Poisson approximation and the
  {Chen-Stein} method,'' \emph{Statistical Science}, vol.~5, no.~4, pp.
  403--424, 1990.

\bibitem{barbour}
A.~Barbour, L.~Holst, and S.~Janson, \emph{Poisson Approximation}.\hskip 1em
  plus 0.5em minus 0.4em\relax Oxford: Clarendon Press, 1992.

\bibitem{SimFW}
\BIBentryALTinterwordspacing
``Proposed simulation framework and theoretical model.'' [Online]. Available:
  \url{https://github.com/andreatassi/SparseRLNC}
\BIBentrySTDinterwordspacing

\end{thebibliography}
\end{document}